# Physics-Informed Neural Operators for Generalizable and Label-Free Inference of Temperature-Dependent Thermoelectric Properties

Hyeonbin Moon[1][†], Songho Lee[1][†], Wabi Demeke[1][†],[*], Byungki Ryu[2], Seunghwa Ryu[1][*]

**Affiliations**

[1]Department of Mechanical Engineering, Korea Advanced Institute of Science and Technology (KAIST), 291 Daehak-ro, Yuseong-gu, Daejeon 34141, Republic of Korea

[2]Energy Conversion Research Center, Korea Electrotechnology Research Institute (KERI), 12 Jeongiui-gil, Seongsan-gu, Changwon-si, Gyeongsangnam-do 51543, Republic of Korea

[†]These authors contributed equally to this work

[*]Corresponding authors: wabi@kaist.ac.kr and ryush@kaist.ac.kr

**Abstract**

Accurate characterization of temperature-dependent thermoelectric properties (TEPs), such as thermal conductivity and the Seebeck coefficient, is essential for reliable modeling and efficient design of thermoelectric devices. However, their nonlinear temperature dependence and coupled transport behavior make both forward simulation and inverse identification difficult, particularly under sparse measurement conditions. In this study, we develop a physics-informed machine learning approach that employs physics-informed neural networks (PINN) for solving forward and inverse problems in thermoelectric systems, and neural operators (PINO) to enable generalization across diverse material systems. The PINN enables field reconstruction and material property inference by embedding governing transport equations into the loss function, while the PINO generalizes this inference capability across diverse materials without retraining. Trained on simulated data for 20 p-type materials and evaluated on 60 unseen materials, the PINO model demonstrates accurate and label-free inference of TEPs using only sparse field data. The proposed framework offers a scalable, generalizable, and data-efficient approach for thermoelectric property identification, paving the way for high-throughput screening and inverse design of advanced thermoelectric materials.

# 1. Introduction

Thermoelectric (TE) materials convert heat directly into electricity via the Seebeck effect, offering a compact and solid-state solution for long-term, low-maintenance energy harvesting in applications ranging from microscale sensors to industrial power systems [1-5]. The performance of a TE device is characterized by the dimensionless figure of merit, $zT = \frac{\sigma \alpha^2}{k} T$, where $\sigma$, $k$, and $\alpha$ represent the electrical conductivity, thermal conductivity, and Seebeck coefficient, respectively [6, 7]. Since these thermoelectric properties (TEPs) are strongly temperature-dependent and coupled, accurate knowledge of their functional forms— $\sigma(T)$, $k(T)$, and $\alpha(T)$ — is critical for material screening, device modeling, and performance prediction. [8-14]

Conventionally, each TEP is measured via separate experimental technique; four-probe setup for $\sigma(T)$, laser-flash thermal diffusivity for $k(T)$, and a steady-state differential method for $\alpha(T)$. Each technique is vulnerable to distinct sources of error, including contact resistance, geometric imperfections, and thermal inhomogeneity [15-17]. These independent uncertainties can accumulate, leading to deviations in estimated $zT$ exceeding 15%, which limits reproducibility and scalability, especially for high-throughput screening and inverse design [18]. A unified inference scheme capable of extracting all three properties from limited measurements while enforcing physical laws would significantly mitigate these challenges.

Recent advances in physics-informed machine learning (PIML) offer a promising alternative for inferring material properties while embedding physical constraints [19-21]. Physics-informed neural networks (PINN) incorporate governing partial differential equations (PDEs), such as the coupled heat and charge transport equations, into their training loss, enabling field reconstruction with minimal labeled data [22-26]. Physics-informed neural operators (PINO) extend this concept further by learning mappings between function spaces, allowing generalization across varying inputs, boundary conditions, and material systems [27-29]. PINN and PINO have been successfully applied to domains such as multiphase transport, solid mechanics, and inverse identification of material or geometric properties [30-33].

Here, we present the first physics-informed learning framework that addresses both forward and inverse problems in coupled thermoelectric systems. First, we train a PINN to solve the coupled PDEs governing heat and charge transport, predicting spatial distributions of temperature and voltage from known TEPs (**Fig. 1(a)**). Next, we adapt the PINN for inverse modeling to infer unknown temperature-dependent Seebeck coefficient and thermal conductivity from sparse temperature and voltage measurements (**Fig. 1(b)**). Finally, we extend this inverse framework using a PINO, enabling inference of temperature-dependent TEPs for previously unseen materials without retraining (**Fig. 1(c)**). Validation on simulated data based on experimentally reported p-type material properties shows less than 2 % error for the forward task and accurate recovery of $\alpha(T)$ and $k(T)$ from fewer than ten interior sensors; the PINO, trained on 20 materials, accurately infers properties for 60 unseen materials. These results demonstrate a viable pathway toward high-throughput, physics-consistent characterization of thermoelectric materials.

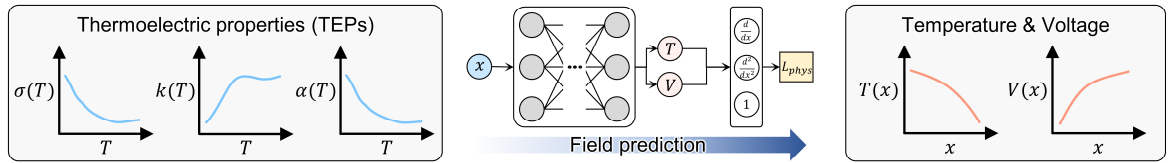

a) PINN for **forward** problem

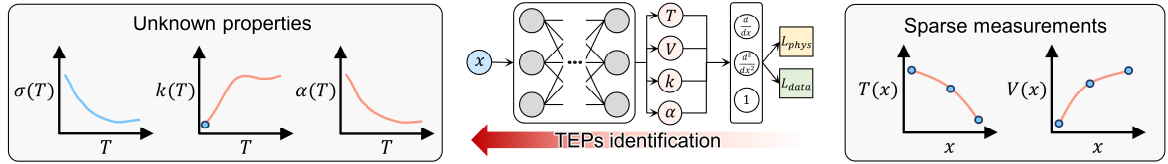

b) PINN for **inverse** problem

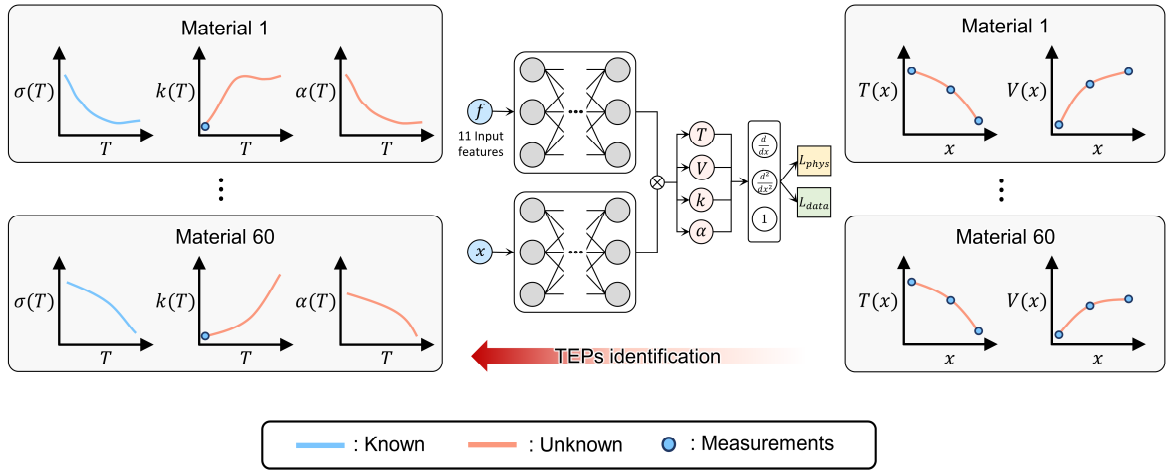

c) PINO for **inverse** problem

**Fig. 1. Schematic illustrations of PINN and PINO frameworks.** (a) Forward problem: PINN predicts temperature and voltage from known thermoelectric properties. (b) Inverse problem: PINN infers unknown properties from sparse measurements. (c) PINO extends inverse modeling to multiple materials using sparse measurements.

## 2. Problem formulation

The steady-state behavior of thermoelectric (TE) materials is governed by the conservation of electric charge and energy [34]:

$$\nabla \cdot \boldsymbol{J} = 0 \tag{1a}$$

$$\nabla \cdot \boldsymbol{q} = \boldsymbol{E} \cdot \boldsymbol{J} \tag{1b}$$

Here, $\boldsymbol{J}$ and $\boldsymbol{q}$ are the electric current density and heat flux, and the electric field is defined as $\boldsymbol{E} = -\nabla V$, where $V$ denotes the voltage (i.e., electric potential). These fluxes depend on the spatial gradients of $V$ and temperature $T$, given by:

$$\boldsymbol{J} = -\sigma(T)(\nabla V + \alpha(T)\nabla T) \tag{1c}$$

$$\boldsymbol{q} = \alpha(T)T\boldsymbol{J} - k(T)\nabla T \tag{1d}$$

where $\sigma(T)$, $k(T)$ and $\alpha(T)$ denote the temperature-dependent electrical conductivity, thermal conductivity, and Seebeck coefficient, respectively.

In many practical thermoelectric applications, thermoelectric generators (TEGs) are designed as 3D modules consisting of multiple p–n legs connected electrically in series and thermally in parallel, as shown in **Fig. 2(a)**. These modules are typically constructed from a single thermoelectric couple, illustrated in **Fig. 2(b)**. Due to the much higher electrical conductivity of the electrodes compared to the thermoelectric material, heat and charge transport are effectively constrained along the primary axis between the hot and cold sides. This directional dominance in transport behavior supports the use of a one-dimensional (1D) approximation. This assumption, validated in **Supplementary Note 1** and supported by previous studies [35, 36], reduces the governing equations to:

$$\frac{dV}{dx} + \alpha(T)\frac{dT}{dx} + \frac{J}{\sigma(T)} = 0 \tag{2a}$$

$$-\frac{d}{dx}\left(k(T)\frac{dT}{dx}\right) + T\frac{d\alpha}{dT}\frac{dT}{dx}J + \frac{J^2}{\sigma(T)} = 0 \tag{2b}$$

Despite the simplification, Eqs. (2a) and (2b) retain their nonlinear and coupled nature due to the strong temperature dependence of the material properties $\sigma(T)$, $k(T)$ and $\alpha(T)$. We consider two distinct but related problems based on the governing equations:

- **Forward problem**: Given the temperature-dependent material properties $\sigma(T)$, $k(T)$ and $\alpha(T)$, along with prescribed boundary conditions, the goal is to compute the spatial distributions of $T(x)$ and $V(x)$. These distributions are essential for evaluating the input thermal flux and output electrical power of TEGs.

- **Inverse problem**: Given sparse point-wise measurements of $T(x)$ and $V(x)$, along with prescribed boundary conditions, the goal is to infer the unknown temperature-dependent material properties, specifically $k(T)$ and $\alpha(T)$. This setup reflects practical experimental conditions, in which only limited field data are available, making conventional numerical approaches inapplicable.

Both problems are defined over a 1D domain of length $L = 100mm$ The hot end $(x = 0)$ is fixed at $T_h = 650K$ with grounded voltage $V = 0$, while the cold end $(x = L)$ is set to $T_c = 350K$, with an applied current density of $J_{applied} = 5000 A/m^2$. In this study, we focus on evaluating spatial field variables and identifying temperature-dependent material properties within a single TEG. This simplified configuration, illustrated in **Fig. 2(c)**, captures the essential transport behavior and reflects the practical conditions under which thermoelectric properties are typically characterized in TEG applications.

The forward problem involves computing the spatial profiles of temperature $T(x)$ and voltage $V(x)$ based on known temperature-dependent material properties $\sigma(T)$, $k(T)$ and $\alpha(T)$. In contrast, the inverse problem, which seeks to identify temperature-dependent TEPs from sparse measurements of $T(x)$ and $V(x)$. However, simultaneously estimating all three properties directly from the measurements results in an ill-posed problem, as multiple combinations of $\sigma(T)$, $k(T)$ and $\alpha(T)$ can produce identical field responses. To ensure well-posedness, we impose a constrained inverse formulation in which $\sigma(T)$ is given, and the thermal conductivity $k(T_0)$ at a reference temperature $T_0 = 350K$ is specified. These constraints balance the number of unknowns with the available measurements, removing ambiguity and enabling stable and physically meaningful inference. Further details on the necessity of this formulation due to ill-posedness are given in **Supplementary Note 2**.

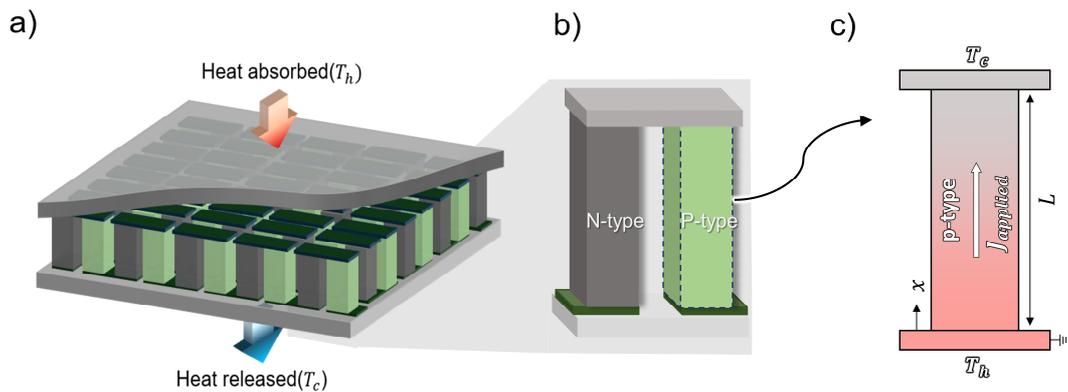

**Fig. 2. Problem formulation for thermoelectric modeling.** (a) A 3D TEG module with multiple p–n pairs. (b) A single p-n couple. (c) 1D single-leg setup.

## 3. Methodology

### 3.1 Physics-informed neural network (PINN)

To solve the forward and inverse thermoelectric problems described in **Section 2**, we first implement a PINN framework. Unlike purely data-driven models, PINN incorporates the governing physical laws, expressed here as 1D steady-state equations, into the training loss function, enabling the model to produce physically consistent solutions even under sparse data conditions.

In the forward problem, illustrated in **Fig. 3(a)**, the forward PINN takes the spatial coordinate $x$ as input and outputs the corresponding temperature $T(x)$ and electric potential $V(x)$. Based on these predictions, the electric current density $J$ and heat flux $q$ are computed using the known material properties $\sigma(T)$, $k(T)$ and $\alpha(T)$. The physical consistency of the predicted fields is enforced by minimizing the residuals of the one-dimensional governing equations (Eqs. (2a) and (2b)), resulting in the total loss:

$$L_{total} = L_{physics} + L_{bc} \tag{3a}$$

$$L_{physics} = \lambda_1 \left\| \frac{dV}{dx} + T\frac{dT}{dx} + \frac{J}{\sigma} \right\|^2 + \lambda_2 \left\| -\frac{d}{dx}\left(k\frac{dT}{dx}\right) + T\frac{d\alpha}{dT}\frac{dT}{dx}J + \frac{J^2}{\sigma} \right\|^2 \tag{3b}$$

$$L_{bc} = \lambda_3 \left\| J|_{x=L} - J_{applied} \right\|^2 \tag{3c}$$

where $\lambda_1, \lambda_2$ and $\lambda_3$ are weights for each loss term. Boundary conditions are enforced via hard constraints to guarantee exact satisfaction:

$$T(x) = x(L-x) \cdot NN_T(x) + T_h + \frac{T_c - T_h}{L}x \tag{4a}$$

$$V(x) = x \cdot NN_V(x) \tag{4b}$$

where $NN_T(x)$ and $NN_V(x)$ denote the unconstrained outputs of the neural network. These formulations ensure that $T|_{x=0} = T_h, T|_{x=L} = T_c$ and $V_{x=0} = 0$ are satisfied exactly. By explicitly encoding the boundary behavior into the functional form, this approach decouples boundary enforcement from the trainable network and eliminates the need for penalty-based loss terms, thereby improving convergence stability and ensuring strict physical fidelity.

For the inverse problem shown in **Fig. 3(b)**, the PINN architecture is extended to infer the unknown temperature-dependent material properties $k(T)$ and $\alpha(T)$ from sparse field measurements. The framework consists of two neural networks: one predicting $T(x)$ and $V(x)$, and another mapping temperature $T$ to the unknown $k(T)$ and $\alpha(T)$. The loss function for inverse problem incorporates an additional data loss term to account for the sparse measurements:

$$L_{total} = L_{physics} + L_{bc} + L_{data} \tag{5a}$$

$$L_{data} = \lambda_4 \|V - V_{measured}\|^2 + \lambda_5 \|T - T_{measured}\|^2 \tag{5b}$$

where $T_{measured}$ and $V_{measured}$ are the sparse sensor measurements, and $\lambda_4$ and $\lambda_5$ are weighting coefficients. Furthermore, an additional constraint is imposed on the network output for $k(T)$ as follow:

$$k(T) = (T - T_0) \cdot NN_k(T) + k(T_0) \tag{6}$$

where $NN_\kappa(T)$ is the unconstratined network output, ensuring that $k(T_0)$ exactly satisfies the known value at $T_0 = 350K$.

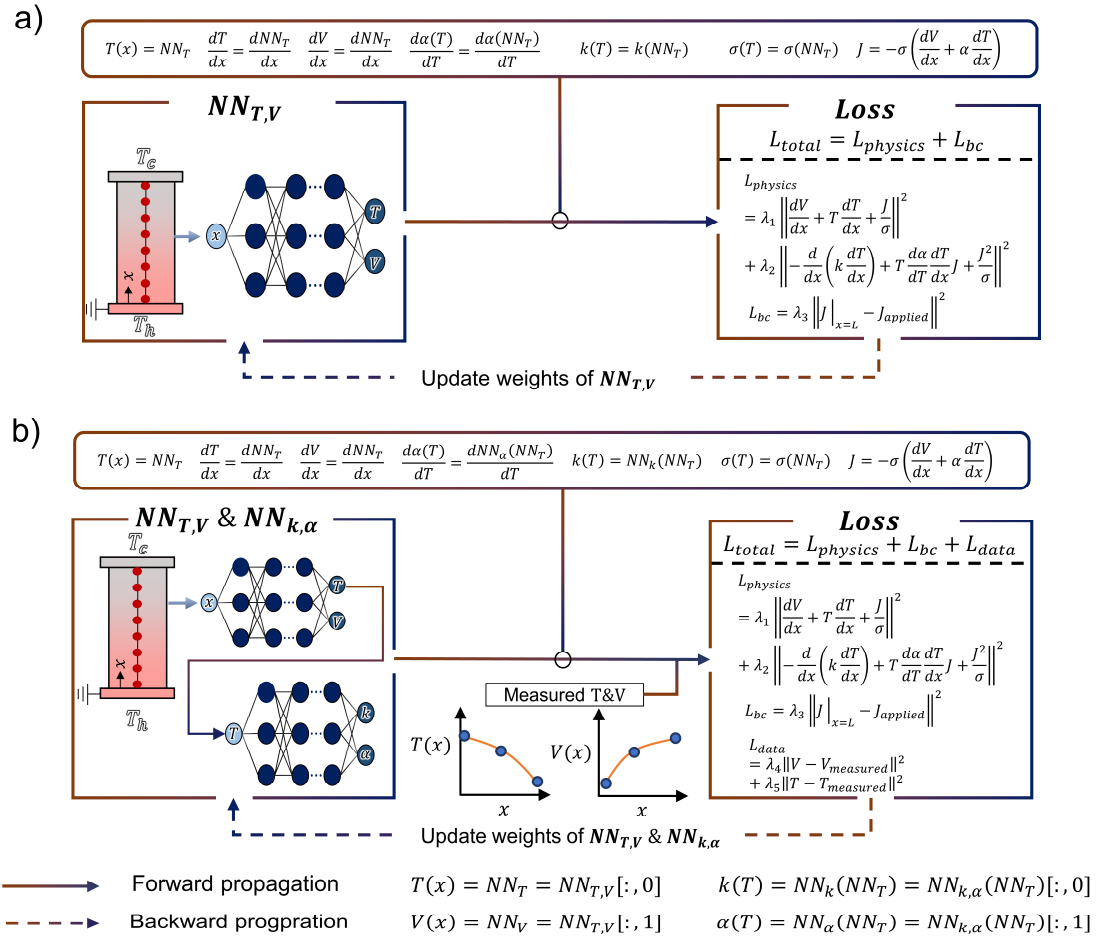

**Fig. 3. PINN-based forward and inverse frameworks.** (a) Forward problem: the PINN predicts spatial temperature and voltage fields from known thermoelectric properties. (b) Inverse problem: the PINN infers unknown temperature-dependent thermoelectric properties from sparse temperature and voltage measurements.

## 3.2 Physics informed neural operator (PINO)

Although PINN effectively solve forward and inverse thermoelectric problems, they are inherently limited to single-instance learning. For materials with different temperature-dependent properties, a conventional PINN must be retrained from scratch, which is impractical for real-world applications involving diverse materials. To overcome this limitation, a PINO framework is adopted, focusing on the inverse problem. PINO leverages the interpolation capability of operator learning to infer the temperature-dependent properties $k(T)$ and $\alpha(T)$ for unseen materials without requiring retraining. This approach enables efficient and scalable identification across a wide range of material types.

The architecture of PINO closely follows the PINN structure used for the inverse problem but replaces each network with a DeepONet architecture comprising branch and trunk networks, as depicted in **Fig 4**. The input to the branch networks includes three sparse temperature measurements, three sparse voltage measurements, four polynomial coefficients representing the temperature-dependent electrical conductivity $\sigma(T)$ (approximated as a third-order polynomial), and a known thermal conductivity value $k(T_0)$, resulting in 11 input features. The trunk network takes either spatial coordinate $x$ or the inferred temperature $T$ as input, depending on the whether the network predicts $T(x)$ and $V(x)$ or material properties $k(T)$ and $\alpha(T)$.

Training of the PINO framework involves optimizing the networks across multiple material cases simultaneously. The total loss function is extended from the PINN formulation and is defined as:

$$L_{total} = L_{physics} + L_{bc} + L_{data} \qquad (7a)$$

where

$$L_{physics} = \lambda_1 \sum_{i=1}^{N_m} \left\| \frac{dV^{(i)}}{dx} + T^{(i)} \frac{dT^{(i)}}{dx} + \frac{J^{(i)}}{\sigma^{(i)}} \right\|^2 +$$

$$\lambda_2 \sum_{i=1}^{N_m} \left\| -\frac{d}{dx}\left( k^{(i)} \frac{dT^{(i)}}{dx} \right) + T^{(i)} \frac{d\alpha^{(i)}}{dT} \frac{dT^{(i)}}{dx} J^{(i)} + \frac{J^{(i)2}}{\sigma^{(i)}} \right\|^2 \quad (7b)$$

$$L_{bc} = \lambda_3 \sum_{i=1}^{N_m} \left\| J^{(i)}|_{x=L} - J^{(i)}_{applied} \right\|^2 \quad (7c)$$

$$L_{data} = \lambda_4 \sum_{i=1}^{N_m} \left\| V^{(i)} - V^{(i)}_{measured} \right\|^2 + \lambda_5 \sum_{i=1}^{N_m} \left\| T^{(i)} - T^{(i)}_{measured} \right\|^2 \quad (7d)$$

Here, $N_m$ denote the number of material cases used during training, and the superscript $(i)$ indicates the corresponding quantities for each case. Furthermore, as in the PINN approach, hard constraints are applied to exactly enforce the boundary conditions for $T(x)$ and $V(x)$ and to satisfy the known thermal conductivity value $k(T_0)$.

More details on the network architectures of the PINN and PINO models used in this study, as well as the input encoding in PINO framework, are provided in **Supplementary Note 3**.

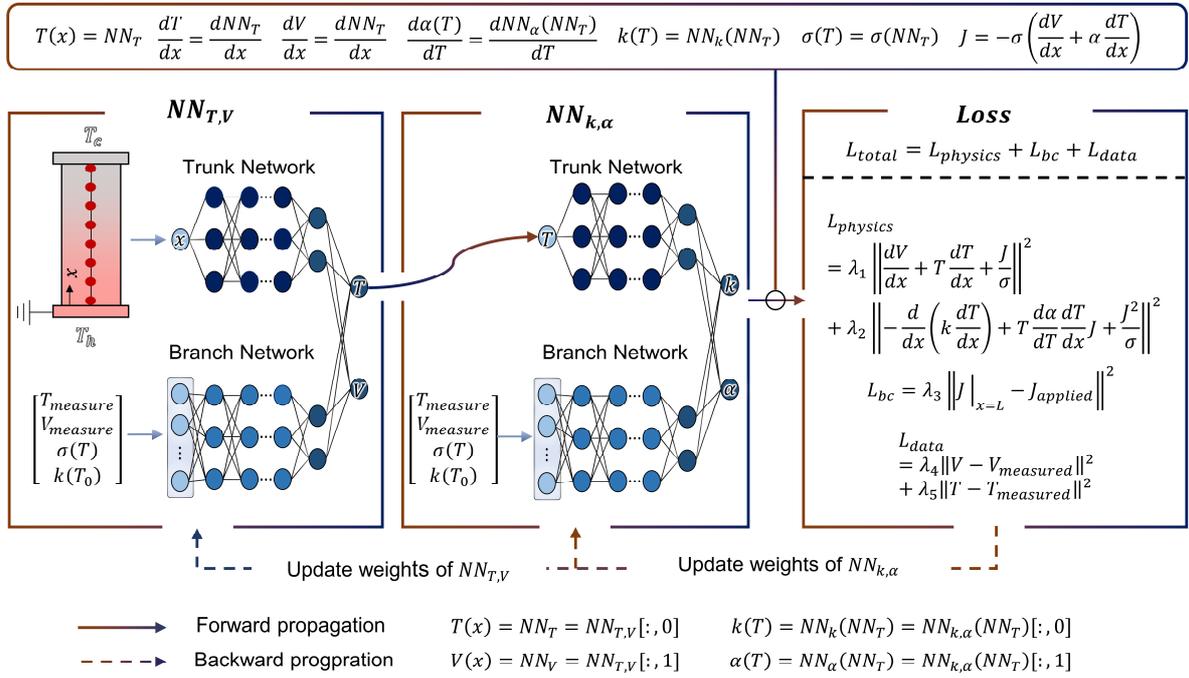

**Fig. 4. PINO framework for inverse problem.** The architecture is composed of two networks, each consisting of a Trunk Network and a Branch Network, making it suitable for handling material variations. The first network, $NN_{T,V}$, is used for field prediction and the second network, $NN_{k,\alpha}$, predicts unknown TEPs. The two networks are trained as composition $NN_{k,\alpha}(NN_{T,V})$, allowing the necessary loss terms to be evaluated.

## 3.3 Non-dimensionalization

Thermoelectric material properties such as electrical conductivity, thermal conductivity, and Seebeck coefficient often differ by several orders of magnitude, which can lead to numerical instability during model training. To enhance numerical stability and enable scale-independent analysis, all governing equations and material properties are expressed in nondimensional form using appropriate reference quantities. Specifically, a reference length $L_{ref}$, temperature $T_{ref}$, electric conductivity $\sigma_{ref}$, thermal conductivity $k_{ref}$ are introduced to normalize quantities as follow:

$$x^* = \frac{x}{L_{ref}}, \quad T^* = \frac{T}{T_{ref}}, \quad k^* = \frac{k}{k_{ref}}, \quad \sigma^* = \frac{\sigma}{\sigma_{ref}}$$

$$\alpha^* = \frac{\alpha}{\sqrt{k_{ref}/(\sigma_{ref} T_{ref})}}, \quad V^* = \frac{V}{\sqrt{(k_{ref} T_{ref})/\sigma_{ref}}}, \quad J^* = \frac{J}{\sqrt{k_{ref} T_{ref} \sigma_{ref}/L_{ref}}}$$

These nondimensional variables are consistently applied throughout the formulation, including the governing equations, boundary conditions, and loss function definitions. The PINN and PINO models are trained using these nondimensional quantities to ensure robust and stable convergence across a range of material properties and scales. All predicted quantities can be readily converted back to their dimensional forms by reapplying the corresponding reference quantities.

## 4. Result and discussion

### 4.1 Data generation

In this study, a database of temperature-dependent TEPs was compiled for 80 p-type materials, based on various studies reported in the literature. The TEP data were digitized using Plot Digitizer software, following the procedure described in our previous work [6]. Detailed references for the selected materials, including digital object identifiers (DOIs), are available in an earlier publication [6, 13, 37]. For each material, the temperature-dependent dependent electrical conductivity $\sigma(T)$, thermal conductivity $k(T)$, and Seebeck coefficient $\alpha(T)$ were interpolated at 25 uniformly spaced points within a common operating temperature range. This range, from $T_c = 350K$ to $T_h = 650K$ was determined based on the overlapping measurement intervals across all materials: specifically, $T_c$ corresponds to the highest reported minimum temperature, and $T_h$ to the lowest reported maximum temperature within the material database. It is worth noting that different materials have different $T_c$ and $T_h$ ranges; however, in this study, we considered the same, $T_c$ and $T_h$ for all materials [13]. The applicability of different available experimental measurements for different materials is not considered in this study for the sake of simplicity. The processed material properties were imported into COMSOL Multiphysics, and finite element simulations were performed under the boundary conditions described in **Section 2** to obtain the spatial profiles of temperature $T(x)$ and voltage $V(x)$ for each material.

Within the PINN framework, a single material was selected from the database as a representative case study (**Fig. 5(a-b)**). In the forward problem setting, the known temperature-dependent properties $\sigma(T)$, $k(T)$ and $\alpha(T)$ were provided as inputs, and the governing equations were solved to predict the spatial profiles of $T(x)$ and $V(x)$. For the corresponding inverse problem, sparse measurements of $T(x)$ and $V(x)$ were synthetically extracted from the COMSOL-simulated fields according to the sampling strategy described in **Section 2**. These sparse data points served as input to the inverse PINN model to reconstruct the underlying thermoelectric properties. The predicted temperature and voltage distributions, along with the reconstructed material properties, were quantitatively compared with the COMSOL simulation results and the original ground truth data to evaluate model performance.

For PINO training and evaluation, 20 materials were randomly selected from the 80-material database. These selected materials were used to construct the training dataset, while the remaining 60 materials served as unseen test cases to evaluate generalization performance (**Fig. 5(c-d)**). Since 20 materials alone provide limited diversity for robust operator learning, a synthetic data augmentation procedure was applied. Specifically, for each selected material, the sparse measurement inputs (three temperature points, three voltage points), the fitted polynomial coefficients for $\sigma(T)$ and the known thermal conductivity $k(T_0)$ were perturbed by up to ±30% using uniform random noise. This augmentation process resulted in approximately 1000 distinct training samples. Further details regarding the data augmentation procedure are provided in **Supplementary Note 4**. It is important to note that both PINN and PINO inverse models were trained in an unsupervised manner, relying solely on physics-based residual losses without using any ground truth values of $k(T)$ and $\alpha(T)$.

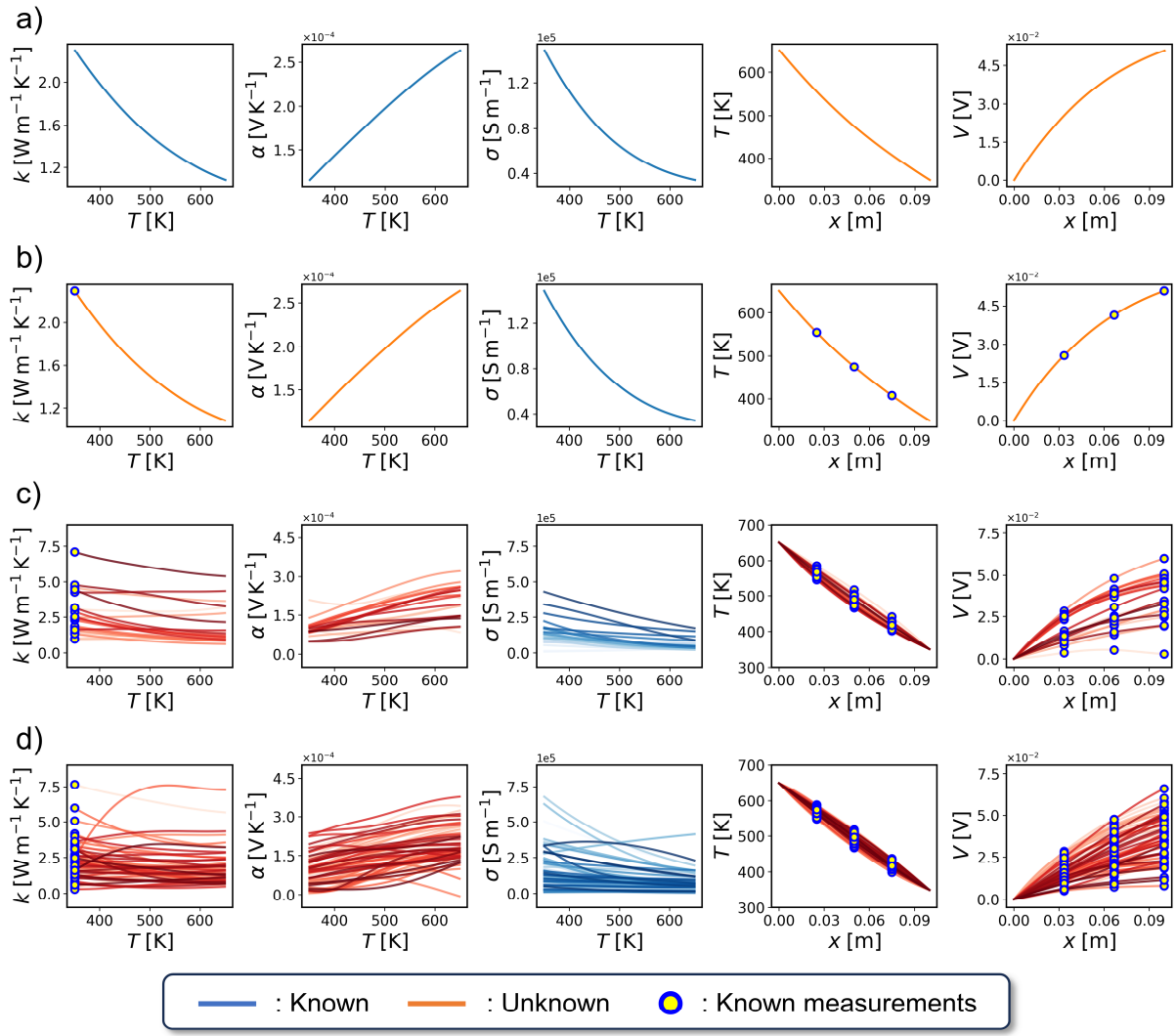

**Fig. 5 Construction of training and test datasets for PINN and PINO.** (a) Ground-truth TEPs used in forward PINN, (b) Sparse measurements extracted for inverse PINN, (c) Training samples for PINO from 20 materials (before augmentation), (d) Test samples for PINO from 60 unseen materials

## 4.2 PINN for forward problem

The PINN framework was applied to the forward thermoelectric problem using the known temperature-dependent material properties $\sigma(T), k(T)$ and $\alpha(T)$, as shown in **Fig. 6(a-c)**. The network was trained by minimizing the residuals of the governing equations and enforcing the boundary conditions. As shown in **Fig. 6(d)**, the total loss, including $L_{physics}$ and $L_{bc}$, converged sufficiently.

The predictive performance of the trained PINN is illustrated in **Figs. 6(e-f)**, where the temperature $T(x)$, voltage $V(x)$, and current density $J(x)$ profiles are compared against the ground truth obtained from COMSOL simulations. The predicted fields show excellent agreement with the reference solutions across the entire domain, with negligible discrepancies. This close match confirms that the PINN successfully learned to solve the coupled thermoelectric governing equations and accurately reproduced the spatial behavior of the physical fields. These results demonstrate that the PINN framework possesses strong capability for solving forward thermoelectric problems involving nonlinear, temperature-dependent material properties.

To further validate the 1D assumption, the gradient fields of $T(x)$ and $V(x)$ obtained from the PINN model are compared with 2D COMSOL simulations, as discussed in **Supplementary Note 1**. The PINN accurately captures the temperature gradient, demonstrating that the model reliably represents the critical aspects of thermoelectric systems. Notably, accurately resolving temperature and voltage gradients along the directions of heat and current flow is essential, particularly for temperature-dependent TE materials [7].

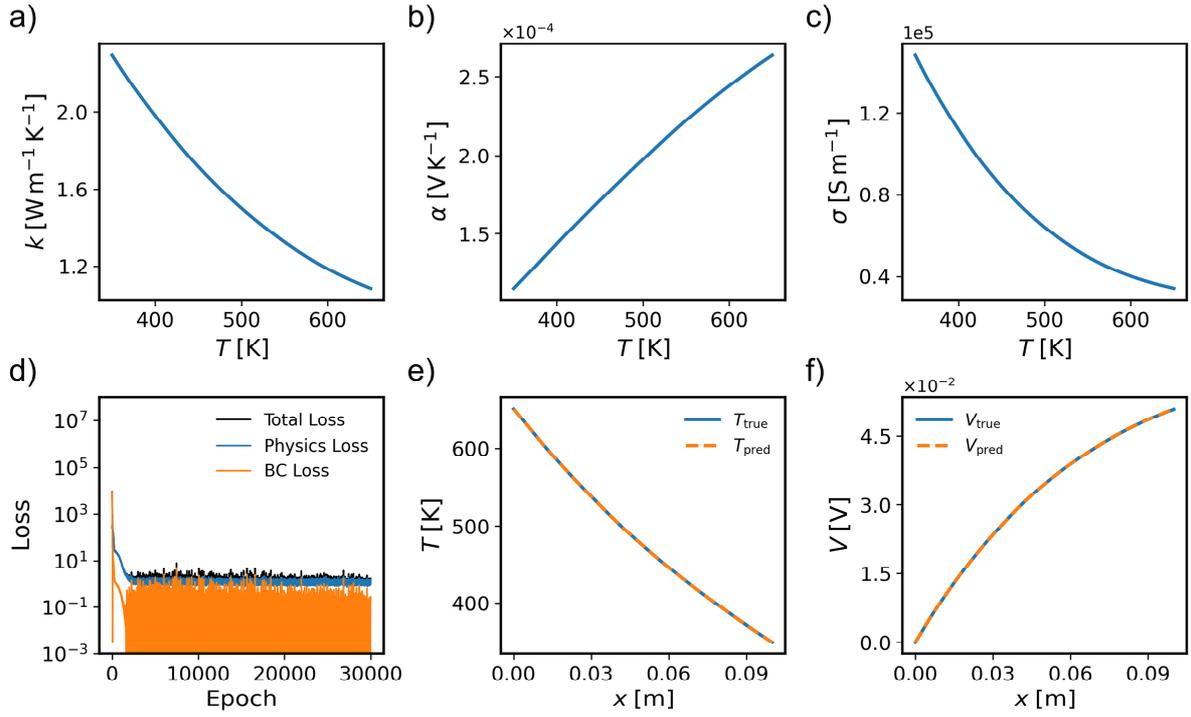

**Fig. 6 PINN results for forward problem.** (a–c) known temperature-dependent material properties, (d) training loss convergence, (e) predicted and true temperature $T(x)$, and (f) predicted and true voltage $V(x)$

### 4.3 PINN for inverse problem

The PINN framework was further applied to the inverse thermoelectric problem, aiming to identify the temperature-dependent material properties $k(T)$ and $\alpha(T)$ from sparse field measurements of temperature $T(x)$ and voltage $V(x)$. In this setting, the electrical conductivity $\sigma(T)$ and a known single value of $k(T_0)$ were assumed to be given, as described in **Section 2**. The network was trained by minimizing a composite loss function comprising PDE residual losses, boundary condition losses, and measurement data losses. As shown in **Fig. 7(a)**, the total loss, including all components, steadily converged during the training process. The predicted temperature $T(x)$, voltage $V(x)$, and current density $J(x)$ profiles are compared with the ground truth in **Figs. 7(b-c)**, showing good agreement across the domain. Furthermore, the identified thermal conductivity $k(T)$ and Seebeck coefficient $\alpha(T)$, presented in **Figs. 7(d-e)**, exhibit excellent agreement with the true properties over the entire temperature range. The known electrical conductivity $\sigma(T)$ used during training is shown in **Fig. 7(f)**. These results demonstrate that the PINN framework can accurately identify key thermoelectric properties from sparse measurement data, validating its effectiveness for inverse modeling in complex Multiphysics systems.

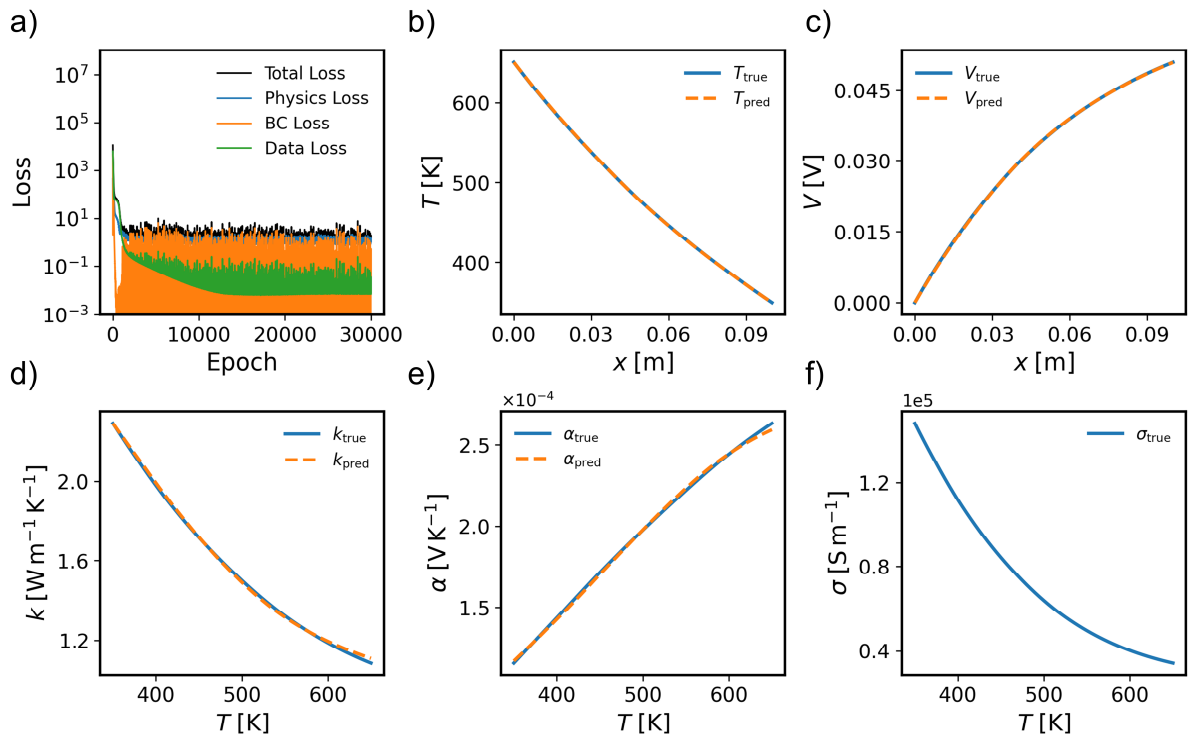

**Fig. 7 PINN results for inverse problem.** (a) loss convergence during training, (b) predicted and true temperature $T(x)$, (c) predicted and true voltage $V(x)$, (d) thermal conductivity $k(T)$, (e) Seebeck coefficient $\alpha(T)$, and (f) known electric conductivity $\sigma(T)$

## 4. PINO for inverse problem

To extend inverse modeling capabilities beyond single-material inference, we developed and trained a physics-informed neural operator (PINO) that generalizes across diverse thermoelectric systems. Unlike the PINN-based approach, which requires retraining for each new material, PINO learns a mapping from sparse field measurements to material properties across multiple materials simultaneously. This transforms the inverse problem from a case-specific formulation into a generalized operator framework, enabling efficient property inference for previously unseen materials without retraining.

As mentioned in Section 4.1, the initial dataset used for training the PINO model was obtained from simulations of 20 distinct thermoelectric materials. To address the limited variability inherent in this relatively small set, we applied data augmentation by introducing perturbations to the input features, including sparse temperature and voltage measurements, polynomial coefficients representing $\sigma(T)$, and a reference thermal conductivity value $k(T_0)$ (see **Supplementary Note 4**). To evaluate the benefit of this augmentation strategy, we trained two versions of the PINO model: one using the augmented dataset and the other using only the original dataset without augmentation. The performance of both models was then compared on 60 previously unseen thermoelectric materials to assess the effect of data augmentation on generalization.

As shown in **Fig. 8(a)**, the training loss of the PINO model converges, indicating stable optimization. The remaining results in **Figs. 8(b–f)** correspond to inference on previously unseen materials. The predicted temperature $T(x)$ and voltage $V(x)$ distributions for a representative unseen material (**Figs. 8(b–c)**) closely match the reference solutions. The inferred thermal conductivity $k(T)$ and Seebeck coefficient $\alpha(T)$, shown in **Figs. 8(d–e)**, align well with true profiles when the model is trained with data augmentation; in contrast, performance degrades without it. The electrical conductivity $\sigma(T)$ used during training is shown in **Fig. 8(f)** for reference.

To assess generalization, the model was evaluated on 60 unseen materials. The results are summarized in **Fig. 9**, where predicted and true values of $k(T)$ and $\alpha(T)$, across all test cases are compared using scatter plots, providing quantitative validation of the trends observed in **Fig. 8**. Both augmented and non-augmented models achieve high accuracy on the training set, indicating that PINO captures the underlying physics effectively. The augmented model shows slightly lower $R^2$ values on training data (0.998 for $k(T)$, 0.981 for $\alpha(T)$) compared to the non-augmented model (0.999 and 0.987, respectively), reflecting the added variability from augmented inputs rather than underfitting. Crucially, the augmented model generalizes far better to unseen materials, achieving $R^2$ values of 0.991 for $k(T)$ and 0.976 for $\alpha(T)$ on the test set—substantially outperforming the non-augmented model, which achieves only 0.638 and 0.900, respectively (**Figs. 9(a–b), 9(d–e)**). These results demonstrate that training on a diverse, augmented dataset improves both robustness and generalization of the PINO framework. Importantly, PINO is trained in an unsupervised manner, without any ground truth labels for $k(T)$ and $\alpha(T)$, relying solely on physics-based residuals from governing equations. Once trained, PINO enables rapid inference of thermoelectric properties for new materials—without retraining. While initial training requires sparse field measurements from only 20 materials, we further validated that this physics-informed unsupervised approach outperforms a fully supervised model trained directly on ground truth properties. A detailed comparison is provided in **Supplementary Note 5**, underscoring the benefit of combining data augmentation with physics-informed learning—especially when limited labeled data are available.

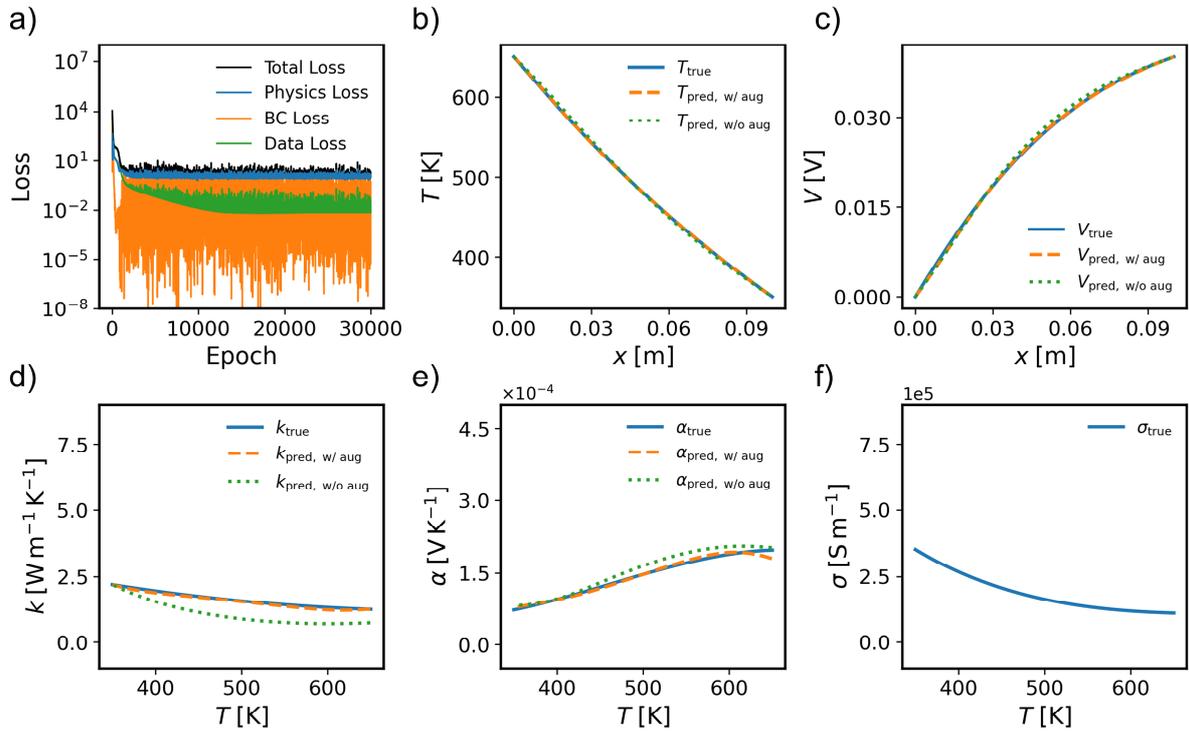

**Fig. 8 PINO results on the inverse problem with and without data augmentation.** (a) Training loss for total, physics, boundary, and data losses. (b,c) Predicted and true temperature $T(x)$ and voltage $V(x)$. (d,e) Inferred thermal conductivity $k(T)$ and Seebeck coefficient $\alpha(T)$ compared to ground truth. (f) Prescribed electrical conductivity $\sigma(T)$, provided as input for inferring other material properties.

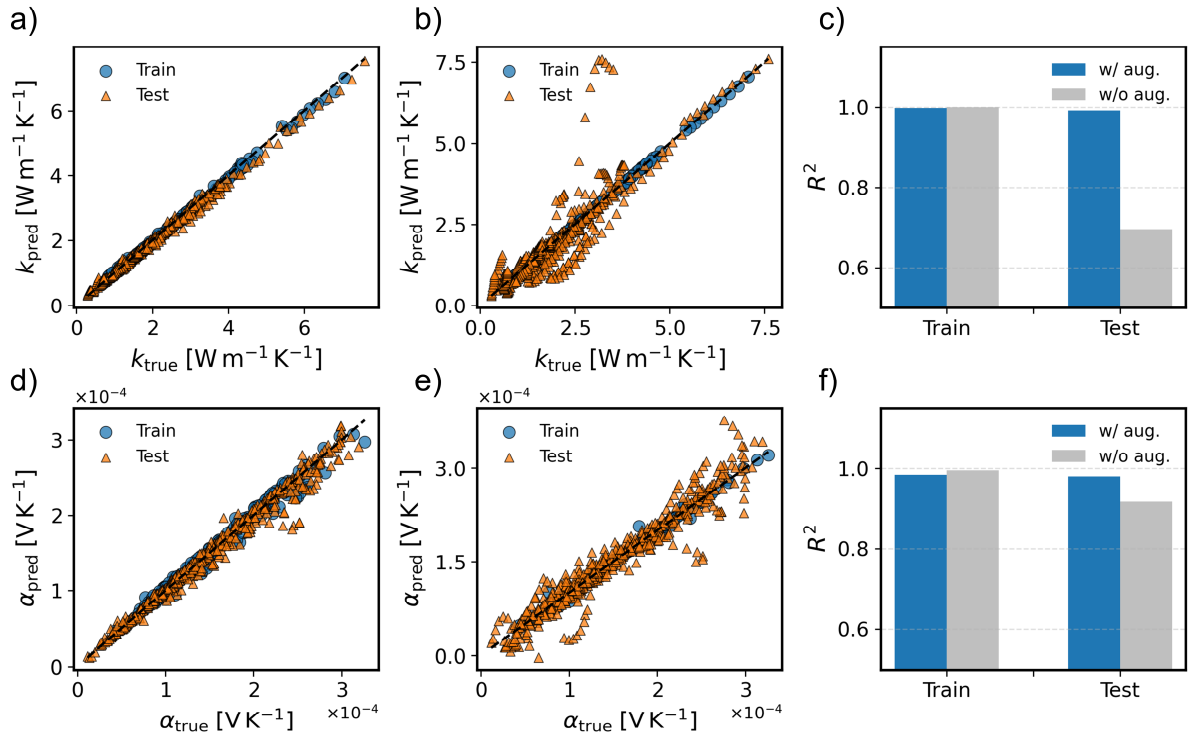

**Fig. 9 Comparison of PINO performance with and without data augmentation.** Predicted versus true values of thermal conductivity $k(T)$ and Seebeck coefficient $\alpha(T)$ with augmentation (a, d) and without augmentation (b, e). The corresponding $R^2$ scores for the training and test datasets are shown in (c) and (f), respectively.

## 5. Conclusion

In this study, we developed a physics-informed machine learning framework that employs PINN for solving forward and inverse problems and PINO for generalizing inverse inference across diverse thermoelectric materials in one-dimensional thermoelectric generators (TEGs) with temperature-dependent material properties. To the best of our knowledge, this work represents one of the first systematic applications of physics-informed learning to coupled thermoelectric systems, incorporating nonlinear thermal–electrical interactions into data-efficient, physically consistent computational models.

First, we constructed a forward PINN that accurately solves the coupled governing differential equations to predict spatial profiles of temperature and voltage, $T(x)$ and $V(x)$, in materials with strongly temperature-dependent properties. Second, we extended the PINN to inverse modeling, enabling the identification of the Seebeck coefficient $\alpha(T)$ and thermal conductivity $k(T)$ from sparse field measurements, using only known electrical conductivity $\sigma(T)$ and a single reference value $k(T_0)$, without requiring labeled property data. Third, we introduced a novel PINO framework that generalizes inverse identification across different material systems. Unlike conventional PINN, the PINO enables on-demand prediction of $\alpha(T)$ and $k(T)$ for previously unseen materials, without retraining, and exhibits strong generalization performance when trained with data augmentation.

While the proposed approach shows high accuracy and scalability, several limitations remain. The current formulation assumes a uniform operating temperature range across all materials, relies on pre-specified $\sigma(T)$, and does not yet capture three-dimensional effects or anisotropic material behavior. These constraints may limit the direct applicability to real-world TEG modules under complex loading or geometry. Future work will aim to extend this framework to multi-dimensional thermoelectric systems, incorporate heterogeneous and anisotropic material properties, and improve robustness to measurement noise and experimental uncertainty. Ultimately, this approach paves the way toward scalable, physics-informed inverse modeling tools for efficient material characterization and design in practical thermoelectric energy conversion applications.

**Data availability**

The thermoelectric materials database used in this work is available at:Ryu B. teMatDb: Thermoelectric Materials Database. GitHub 2025. https://github.com/byungkiryu/teMatDb (accessed May 22, 2025)..

**Code availability**

The codes developed in this study are available from the authors upon reasonable request.

**Corresponding authors**

Correspondence to Seunghwa Ryu and Wabi Demeke

**Acknowledgements**

This work was supported by the National Research Foundation of Korea(NRF) grand funded by the Korea government(MSIT) (No. RS-2023-00222166 and No. RS-2023-00247245).

**Author contributions**

**Hyeonbin Moon:** Conceptualization, Methodology, Investigation, Data curation, Software, Writing - original draft, Writing - review & editing.

**Songho Lee:** Conceptualization, Methodology, Investigation, Data curation, Software, Writing - original draft, Writing - review & editing.

**Wabi Demeke:** Conceptualization, Validation, Data curation, Investigation, Software, Supervision, Writing - original draft, Writing - review & editing.

**Byungki Ryu:** Data curation, Investigation, Discussion, Writing - review & editing.

**Seunghwa Ryu:**   Formal analysis, Validation, Supervision, Project administration, Writing – review & editing.

**Competing interests**

The authors declare that they have no known competing financial interests or personal relationships that could have appeared to influence the work reported in this paper

**References**

1. Gao, W., Y. Wang, and F. Lai, Thermoelectric energy harvesting for personalized healthcare. Smart medicine, 2022. **1**(1): p. e20220016.

2. Jaziri, N., et al., A comprehensive review of Thermoelectric Generators: Technologies and common applications. Energy reports, 2020. **6**: p. 264-287.

3. Liu, Y., et al., Integrated micro thermoelectric devices with self-power supply and temperature monitoring: Design and application in power grid early warning. Applied Thermal Engineering, 2024. **247**: p. 122922.

4. Rao, Y., T. Bechtold, and D. Hohlfeld, Design of packaged thermoelectric generators for implantable medical devices: A comprehensive parameter study. Energy, 2025. **320**: p. 134932.

5. Xu, Z., et al., Waste heat recovery of power plant with large scale serial absorption heat pumps. Energy, 2018. **165**: p. 1097-1105.

6. Ryu, B., J. Chung, and S. Park, Thermoelectric degrees of freedom determining thermoelectric efficiency. Iscience, 2021. **24**(9).

7. Ryu, B., J. Chung, and S. Park, Thermoelectric algebra made simple for thermoelectric generator module performance prediction under constant Seebeck coefficient approximation. Journal of Applied Physics, 2025. **137**(5).

8. Biswas, K., et al., High-performance bulk thermoelectrics with all-scale hierarchical architectures. Nature, 2012. **489**(7416): p. 414-418.

9. Kim, S.I., et al., Dense dislocation arrays embedded in grain boundaries for high-performance bulk thermoelectrics. Science, 2015. **348**(6230): p. 109-114.

10. Zhao, L.-D., et al., Ultralow thermal conductivity and high thermoelectric figure of merit in SnSe crystals. nature, 2014. **508**(7496): p. 373-377.

11. Zhou, C., et al., Polycrystalline SnSe with a thermoelectric figure of merit greater than the single crystal. Nature materials, 2021. **20**(10): p. 1378-1384.

12. Mao, J., et al., High thermoelectric cooling performance of n-type Mg3Bi2-based materials. Science, 2019. **365**(6452): p. 495-498.

13. Demeke, W., B. Ryu, and S. Ryu, Machine learning-based optimization of segmented thermoelectric power generators using temperature-dependent performance properties. Applied Energy, 2024. **355**: p. 122216.

14. Zhang, L. and F. Liu, High-throughput approach to explore cold metals for electronic and thermoelectric devices. npj Computational Materials, 2024. **10**(1): p. 78.

15. Borup, K.A., et al., Measuring thermoelectric transport properties of materials. Energy & Environmental Science, 2015. **8**(2): p. 423-435.

16. Mavrokefalos, A., et al., Four-probe measurements of the in-plane thermoelectric properties of nanofilms. Review of scientific instruments, 2007. **78**(3).

17. Cha, J., J. Seo, and S. Kim, Building materials thermal conductivity measurement and correlation with heat flow meter, laser flash analysis and TCi. Journal of thermal analysis and calorimetry, 2012. **109**(1): p. 295-300.

18. Wei, T.-R., et al., How to measure thermoelectric properties reliably. Joule, 2018. **2**(11): p. 2183-2188.

19. Karniadakis, G.E., et al., Physics-informed machine learning. Nature Reviews Physics, 2021. **3**(6): p. 422-440.

20. Cuomo, S., et al., Scientific machine learning through physics–informed neural networks: Where we are and what's next. Journal of Scientific Computing, 2022. **92**(3): p. 88.

21. Meng, C., et al., When physics meets machine learning: A survey of physics-informed machine learning. Machine Learning for Computational Science and Engineering, 2025. **1**(1): p. 1-23.

22. Khatamsaz, D., et al., A physics informed bayesian optimization approach for material design: application to NiTi shape memory alloys. npj Computational Materials, 2023. **9**(1): p. 221.

23. Li, R., et al., Physics-informed deep learning for solving phonon Boltzmann transport equation with large temperature non-equilibrium. npj Computational Materials, 2022. **8**(1): p. 29.

24. Zhou, J., R. Li, and T. Luo, Physics-informed neural networks for solving time-dependent mode-resolved phonon Boltzmann transport equation. npj Computational Materials, 2023. **9**(1): p. 212.

25. Chen, Z., Y. Liu, and H. Sun, Physics-informed learning of governing equations from scarce data. Nature communications, 2021. **12**(1): p. 6136.

26. Akhare, D., et al., Probabilistic physics-integrated neural differentiable modeling for isothermal chemical vapor infiltration process. npj Computational Materials, 2024. **10**(1): p. 120.

27. Lu, L., P. Jin, and G.E. Karniadakis, Deeponet: Learning nonlinear operators for identifying differential equations based on the universal approximation theorem of operators. arXiv preprint arXiv:1910.03193, 2019.

28. Li, Z., et al., Physics-informed neural operator for learning partial differential equations. ACM/JMS Journal of Data Science, 2024. **1**(3): p. 1-27.

29. Oommen, V., et al., Rethinking materials simulations: Blending direct numerical simulations with neural operators. npj Computational Materials, 2024. **10**(1): p. 145.

30. Moon, H., et al., Physics-Informed Neural Network-Based Discovery of Hyperelastic Constitutive Models from Extremely Scarce Data. arXiv preprint arXiv:2504.19494, 2025.

31. Diao, Y., et al., Solving multi-material problems in solid mechanics using physics-informed neural networks based on domain decomposition technology. Computer Methods in Applied Mechanics and Engineering, 2023. **413**: p. 116120.

32. Hu, H., L. Qi, and X. Chao, Physics-informed Neural Networks (PINN) for computational solid mechanics: Numerical frameworks and applications. Thin-Walled Structures, 2024: p. 112495.

33. Qiu, R., et al., Physics-informed neural networks for phase-field method in two-phase flow. Physics of Fluids, 2022. **34**(5).

34. Ge, Y., et al., Optimal design of a segmented thermoelectric generator based on three-dimensional numerical simulation and multi-objective genetic algorithm. Energy, 2018. **147**: p. 1060-1069.

35. Chen, B., et al., Experimental verification of one-dimensional models of thermoelectric generators. Physical Review E, 2025. **111**(4): p. 045506.

36. Sheikhnejad, Y., et al., Laser thermal tuning by transient analytical analysis of peltier device. IEEE Photonics Journal, 2017. **9**(3): p. 1-13.

37. Ryu, B., et al., teMatDb: A High-Quality Thermoelectric Material Database with Self-Consistent ZT Filtering. arXiv preprint arXiv:2505.19150, 2025.

# Supplementary Note

# Physics-Informed Neural Operators for Generalizable and Label-Free Inference of Temperature-Dependent Thermoelectric Properties

Hyeonbin Moon[1†], Songho Lee[1†], Wabi Demeke[1†,*], Byungki Ryu[2], Seunghwa Ryu[1]*

**Affiliations**

[1]Department of Mechanical Engineering, Korea Advanced Institute of Science and Technology (KAIST), 291 Daehak-ro, Yuseong-gu, Daejeon 34141, Republic of Korea

[2]Energy Conversion Research Center, Korea Electrotechnology Research Institute (KERI), 12 Jeongiui-gil, Seongsan-gu, Changwon-si, Gyeongsangnam-do 51543, Republic of Korea

[†]These authors contributed equally to this work

*Corresponding authors: wabi@kaist.ac.kr and ryush@kaist.ac.kr

**Supplementary Note 1. Comparison of 1D versus 2D**

While a one-dimensional (1D) assumption was adopted to evaluate the field variables of thermoelectric (TE) materials, we validated the validity of this simplification by comparing it against two-dimensional (2D) simulations. Specifically, the 1D PINN model solution was copied in the y-direction and compared with full 2D finite element analysis (FEA) results. As shown in **Fig. S1(a–c)** for temperature and **Fig. S1(d–f)** for voltage, the predictions from the 1D PINN model closely match the corresponding 2D FEA solutions. This agreement supports the validity of the 1D approximation to higher dimensions. Notably, the critical aspect of solving TE systems lies in accurately capturing the temperature and voltage gradients along the direction of heat and current flow, especially for temperature-dependent TE material properties.

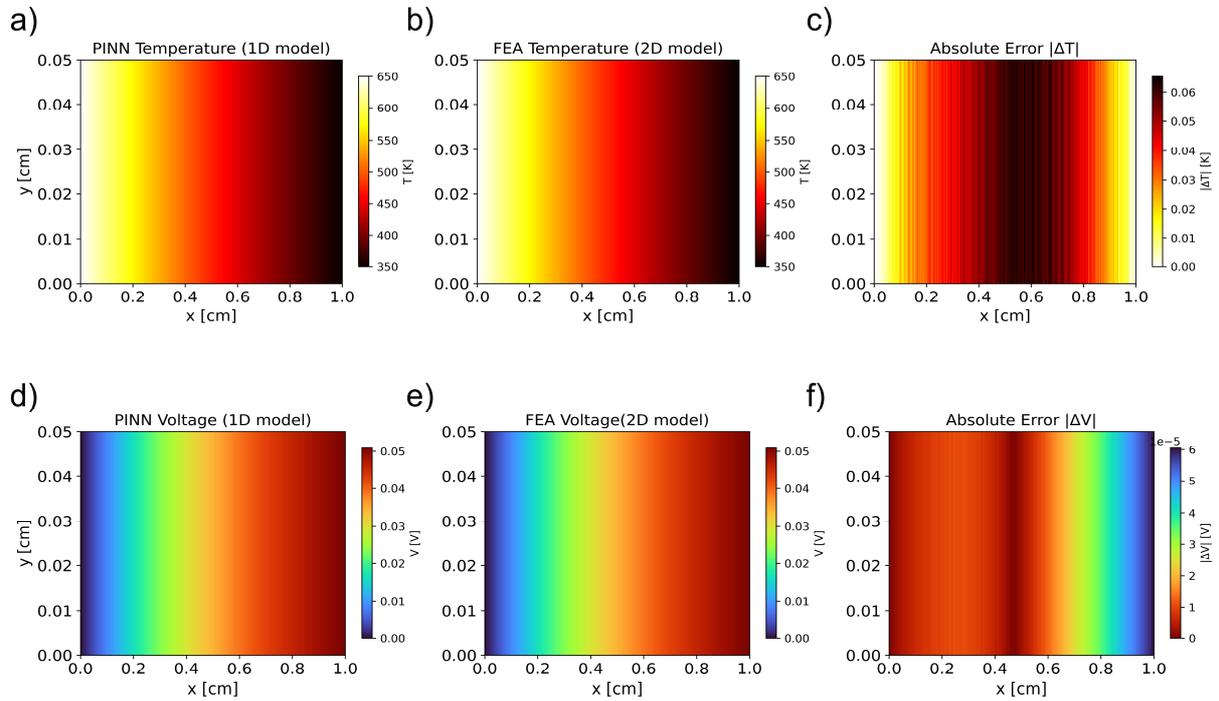

**Fig. S1** (a) Temperature fields obtained by 1-PINN model by simply copying in y-direction. (b) the temperature fields obtained by 2-D analysis of FEA. (c) temperature difference between PINN vs 2-D FEA models. (d) voltage fields obtained by 1-PINN model by simply copying in y-direction. (b) the voltage fields obtained by 2-D analysis of FEA. (c) voltage difference between PINN vs 2-D FEA models.

**Supplementary Note 2. Ill-posedness of inverse problem**

We consider the steady-state thermoelectricity in one dimension, governed by the following coupled differential equations:

$$\frac{d}{dx}\left(-\sigma(T)\frac{dV}{dx} + \alpha(T)\frac{dT}{dx}\right) = 0 \tag{1a}$$

$$\frac{d}{dx}\left(\alpha(T)T\left(-\sigma(T)\frac{dV}{dx} + \alpha(T)\frac{dT}{dx}\right) - k(T)\frac{dT}{dx}\right) = -\frac{dV}{dx}\left(-\sigma(T)\frac{dV}{dx} + \alpha(T)\frac{dT}{dx}\right) \tag{1b}$$

subject to the boundary conditions:

$$T(x=0) = T_h, \quad T(x=L) = T_h \tag{1c}$$

$$V(x=0) = 0, \quad J(x=L) = J_{applied} \tag{1d}$$

Suppose the spatial profiles $T(x)$ and $V(x)$ are fully known from measurements. The inverse problem then aims to identify the three temperature-dependent material properties $\sigma(T)$, $k(T)$ and $\alpha(T)$. However, the direct identification is fundamentally ill-posed, as the system provides only two scalar equations at each spatial point while requiring the estimation of three unknown functions.

To address this underdetermination, two assumptions are introduced. First, the temperature-dependent electrical conductivity $\sigma(T)$ is assumed to be fully known from independent measurements, thereby reducing the number of unknowns from three to two. Second, to supplement the boundary information, it is assumed that the thermal conductivity $k(T)$ is known at a specific reference temperature $T_0$.

Under these assumptions, the inverse problem reduces to determining two unknown functions, $k(T)$ and $\alpha(T)$, based on the two governing differential equations (Eqs. (1a) and (1b)) and two boundary-related constraints: the prescribed current density $J(x=L) = J_{applied}$ and the known value $k(T_0)$. This formulation ensures that the number of independent constraints matches the number of unknowns, thereby leading to a mathematically well-posed inverse problem. The assumptions of known $\sigma(T)$ and $k(T_0)$ arise directly from the mathematical structure of the governing system and are essential for preventing non-uniqueness and ensuring the identifiability of the material properties.

**Supplementary Note 3. PINN/PINO architecture and input encoding**

As shown in **Fig. S2,** all networks are 6-layer multilayer perceptron's (MLPs) with 32 nodes per layer and hyperbolic tangent (Tanh) activation. The forward PINN uses a single network. The inverse PINN uses two networks with identical structure. The inverse PINO consists of four networks—two branches and two trunks—with shared architecture but different inputs. All models were trained using the Adam optimizer with a fixed learning rate of $1 \times 10^{-4}$

For the inverse PINO, each branch network receives an 11-dimensional input vector explicitly defined as

$$\{T_{measure}^1, T_{measure}^2, T_{measure}^3, V_{measure}^1, V_{measure}^2, V_{measure}^3, \sigma_1, \sigma_2, \sigma_3, \sigma_4, k(T_0)\}$$

Here, $T_{measure}^i, V_{measure}^i$ $(i = 1,2,3)$ represent sparse point measurements of temperature and voltage along the spatial domain. The coefficients $\sigma_1$ to $\sigma_4$ correspond to a third-order polynomial approximation of the temperature-dependent electrical conductivity $\sigma(T)$, providing a simplified representation of its variation with temperature. The final term $k(T_0)$ provides a known single value of thermal conductivity. All input features are non-dimensionalized to ensure numerical stability and consistent scaling during training.

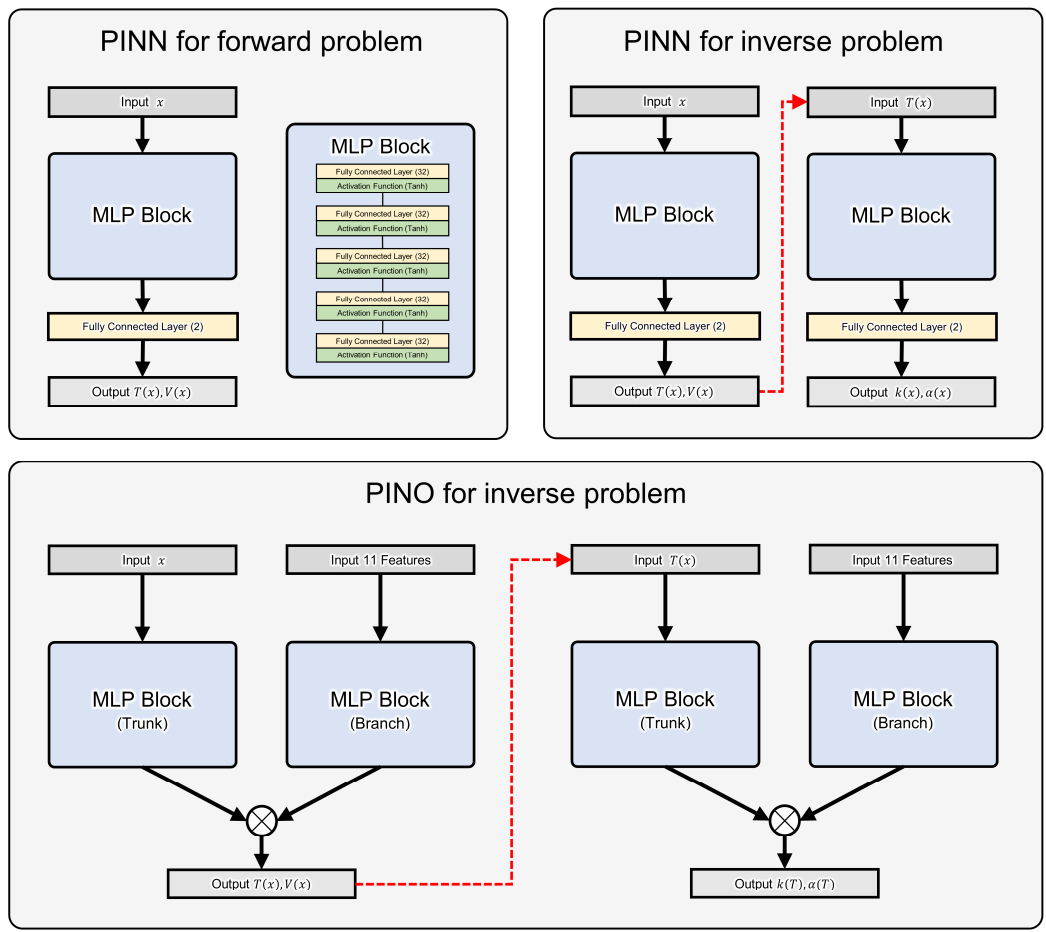

**Fig. S2 Architectures of PINN and PINO used for forward and inverse problems.**

**Supplementary Note 4. Data augmentation for PINO training**

The input features to the branch networks in the PINO framework consist of eleven elements: three sparse temperature measurements $T(x)$, three sparse voltage measurements $V(x)$, four coefficients of a third-order polynomial fit to the electrical conductivity $\sigma(T)$, and a known value of thermal conductivity $k(T_0)$ at $T_0 = 350K$. Thus, each training sample can be represented as a vector in $\mathbf{R}^{11}$. To enable supervised-free learning across multiple material systems, twenty distinct material cases were initially selected from the available dataset, providing a set of base inputs $\{\mathbf{y_i}\}_{i=1}^{20}$, where each $\mathbf{y_i} \in \mathbf{R}^{11}$. These base cases were generated based solely on physical measurements $(T, V, \sigma(T), k(T_0))$ without reference to ground truth thermal conductivity $k(T)$ or Seebeck coefficient $\alpha(T)$ during training.

However, training PINO solely on these 20 samples resulted in insufficient generalization capability. To overcome this, data augmentation was introduced by applying random perturbations to the input features. Specifically, each input feature $y_j$ was independently multiplied by a random scaling factor $r_{ij} \sim Uniform\ (0.7, 1.3)$, leading to augmented inputs of the form:

$$\widehat{y_{ij}} = r_{ij} \times y_{ij} \quad for\ i = 1, \dots, 20\ and\ j = 1, \dots, 11$$

This augmentation process was repeated 49 times, producing approximately 1000 perturbed samples in total for training.

It is important to note that augmentation was performed by perturbing the physically realizable base cases rather than randomly sampling arbitrary feature vectors. Direct random sampling of temperature and voltage measurements, along with $\sigma(T)$ and $k(T_0)$ values, could result in non-physical or inconsistent data that does not correspond to any realizable set of thermoelectric material properties. Such inconsistencies would introduce noise into the training process and deteriorate model performance. By perturbing existing physically meaningful samples, we ensured that all augmented inputs remained within a physically feasible regime, allowing PINO to learn inverse mappings robustly.

**Supplementary Note 5. Comparison with purely data-driven approach**

Typical data-driven approaches rely on labeled datasets during training. While standard PINO does not have access to full $\kappa(T)$ and $\alpha(T)$ curves, it can reconstruct complete temperature-dependent profiles using only sparse field measurements and known electrical conductivity. To highlight the limitations of purely data-driven methods, we trained a multi-layer perceptron (MLP) using sparse measurements from 20 TE materials, including $\sigma(T)$ and a single thermal conductivity value, as inputs to predict the full $\kappa(T)$ and $\alpha(T)$ curves (**Fig. S3(a)**). The model was then tested on 60 unseen materials. As shown in **Fig. S3(b)** and **S3(c)**, the MLP overfits the training data and fails to generalize, demonstrating the limitations of data-driven models when labeled data are limited.

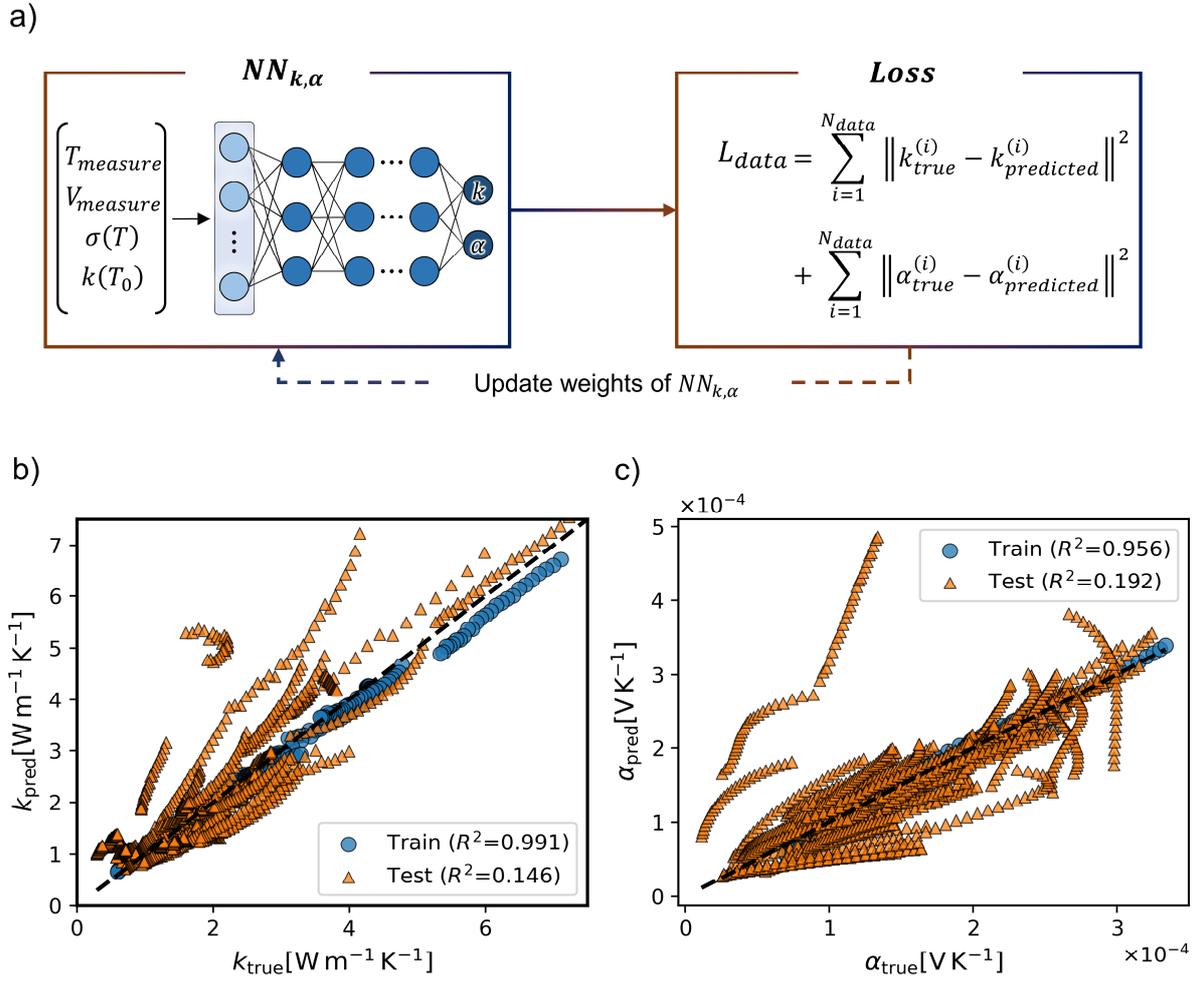

**Fig. S3** (a) Training of the MLP model fully data-driven approach (b) and (c) Comparison between predicted and true values for the training and test sets, for thermal conductivity and Seebeck coefficient, respectively.